\documentclass[12pt]{article}

\csname @addtoreset\endcsname{equation}{section}
\newcommand{\eq}{\begin{equation}}
\newcommand{\feq}{\end{equation}}
\newcommand{\eqn}{\begin{eqnarray}}
\newcommand{\feqn}{\end{eqnarray}}
\newcommand{\arr}{\begin{eqnarray*}}
\newcommand{\farr}{\end{eqnarray*}}

\newcommand{\R}{{\mathbf{R}}}

\newcommand{\Z}{{\mathbf{Z}}}

\def\al{\alpha}
\def\be{\beta}
\def\ga{\gamma}
\def\de{\delta}

\def\om{\omega}

\begin{document}
\begin{titlepage}
\begin{flushright}
IFUM-645-FT \\
UTF-436\\
hep-th/9910065
\end{flushright}
\vspace{.3cm}
\begin{center}
{\Large \bf  $w_{\infty}$ Algebras, Conformal Mechanics, and Black Holes}
\vfill%\vskip 15mm%27.mm
{\large \bf {Sergio Cacciatori$^1$,
Dietmar Klemm$^2$ and
Daniela Zanon$^1$}}\\ 
\vfill%\vskip 7mm%1cm
{\small
$^1$ Dipartimento di Fisica dell'Universit\`a di Milano
and\\ INFN, Sezione di Milano,
Via Celoria 16,
20133 Milano, Italy.\\
\vspace*{0.4cm}
$^2$ Dipartimento di Fisica dell'Universit\`a di Trento
and\\ INFN, Gruppo Collegato di Trento,
Via Sommarive 14,
38050 Povo (TN), Italy.\\}
\end{center}
\vfill
\begin{center}
{\bf Abstract}
\end{center}
{\small 
We discuss BPS solitons in gauged ${\cal N}=2$, $D=4$ supergravity. The
solitons represent extremal black holes interpolating between different vacua
of anti-de Sitter spaces. The isometry superalgebras 
are determined and the motion of a superparticle in the extremal
black hole background is studied and confronted with superconformal
mechanics. We show that the  Virasoro symmetry 
of conformal mechanics, which describes the dynamics of the superparticle
near the horizon of  the extremal black hole under consideration,
extends to a symmetry under the $w_{\infty}$ algebra
of area-preserving diffeomorphisms. 
We find that a Virasoro subalgebra of $w_{\infty}$ can be
associated to the Virasoro algebra of the asymptotic symmetries of $AdS_2$.
In this way spacetime diffeomorphisms of
$AdS_2$ translate into diffeomorphisms in phase space: our
system offers an explicit realization of the $AdS_2/CFT_1$
correspondence. Using the dimensionally reduced action, the central charge
is computed. Finally, we
also present generalizations of superconformal mechanics
which are invariant under ${\cal N} =1$ and ${\cal N} =2$ 
superextensions of $w_{\infty}$.}
\vspace{2mm} \vfill \hrule width 3.cm
\begin{flushleft}
e-mail: cacciatori@mi.infn.it\\
e-mail: klemm@science.unitn.it \\
e-mail: daniela.zanon@mi.infn.it
\end{flushleft}
\end{titlepage}

\section{Introduction}
Solitonic objects play an important role in string theory. In
particular, the study of certain $p$-branes in supergravity theories,
which interpolate between Minkowski space at infinity, and products of
anti-de Sitter spaces and compact Einstein manifolds near
the horizon, led to the $AdS/CFT$ correspondence
\cite{malda,kleb1,witten1}.
In the present paper, we extend the discussion of interpolating solitons
to the case of gauged supergravities. In particular we concentrate on the
gauged ${\cal N}=2$, $D=4$ theory \cite{das,frad}.
A salient feature of the solitons in the gauged
theory, which distinguishes them from corresponding objects in the
ungauged case, is that their near-horizon limit involves not only
products of $AdS$ spaces and positive curvature manifolds
(like the $AdS_2\times S^2$ Bertotti-Robinson solution,
which arises as the near-horizon limit of the extremal Reissner-Nordstr\"om
black hole), but also
spacetimes like $AdS_2\times H^2$
($H^2$ denoting the two-dimensional hyperbolic space).\\
We study the motion of a superparticle in the near-horizon black hole
background  and analyze the associated conformal mechanics model. 
The isometries of the theory give rise to two copies of the
Virasoro algebra, in the angular sector and in the
radial-time sector of the superparticle action respectively.
Focusing on the latter
we show that the corresponding Hamiltonian can be written in the most
general scale-invariant form, namely
\eq
H=\frac{p^2}{2f(u)},
\label{hamilt}
\feq
where $q$, $p$ are the canonical conjugate variables and $f$ is an
arbitrary function of $u=pq$.
The Virasoro symmetry, which exists for any system with scale-invariant
Hamiltonian of one dynamical variable \cite{kumar}, is only a subalgebra of a
larger symmetry
algebra, namely the algebra of certain volume-preserving diffeomorphisms.
It is interesting to observe that the same symmetry algebra
was found in string theories with
two-dimensional target space \cite{aj1,minic,witten,aj2}.
In particular we discuss a
$w_{\infty}$ subalgebra of area-preserving diffeomorphisms\footnote{This
algebra has been encountered previously in the theory of
supermembranes \cite{dewit}, and has also been considered
in the context of two-dimensional
black holes and the matrix model \cite{ellis,jevicki}.}.
This symmetry occurs due to the fact that one can find a canonical
transformation which reduces the scale-invariant Hamiltonian of one
dynamical variable to that of a free particle. Then the generators of
$w_{\infty}$ act as symplectic diffeomorphisms preserving the
two-form $\Omega = dp \wedge dq - dH \wedge dt$. We find that the
algebra admits one central extension (since the first Betti number
of the phase space is $b_1=1$).
For this general class of models we
show that in a natural way
a Virasoro subalgebra of $w_{\infty}$ can be associated to the Virasoro
algebra of  the asymptotic symmetries of $AdS_2$ \cite{cadoni}.
In this way the quantum-mechanical
system explicitly realizes the asymptotic $AdS_2$ symmetries.
The connection
between gravity on $AdS_2$ and conformal field theory in 0+1 dimensions
can be regarded as an example of the
$AdS_2/CFT_1$ correspondence \cite{malda,mms,cadoni,strominger,nakatsu}.
On the gravity (bulk) side, the Virasoro algebra is generated by
spacetime diffeomorphisms preserving the asymptotic form of the metric.
In the theory living on the boundary of $AdS_2$,
these Virasoro generators translate into generators of
diffeomorphisms in the particle phase space. For all black holes
whose near-horizon metric contains an $AdS_2$ factor, a central
charge appears in this Virasoro algebra.
Finally the above mentioned canonical transformation allows us to construct
models of superconformal mechanics exhibiting the symmetries of
${\cal  N} =1$ and ${\cal  N} =2$ superextensions of $w_{\infty}$. 

The remainder of this paper is organized as follows:\\
In section \ref{modsol} we present the model and the BPS interpolating
soliton solutions. We analyze their near-horizon limit and their
supersymmetry properties. Furthermore, the isometry superalgebras of
the soliton and its near-horizon limit are derived.
In section \ref{confmech} the motion of a particle near the horizon
of the solitonic black hole is studied and compared with
conformal mechanics.
The symmetries of conformal mechanics are studied in detail
in section \ref{symm}. In section
\ref{adscft}
we show that the asymptotic symmetries of the bulk theory living 
on $AdS_2$ are in direct correspondence with the symmetries of the
conformal theory living on the boundary.
The supersymmetric extensions are presented in section \ref{super}.
Finally our results are summarized and
discussed.

\section{Interpolating solitons in ${\cal N}=2$,
$D=4$ gauged Supergravity} \label{modsol}

Let us first briefly review
the gauged version of ${\cal N}=2$, $D=4$ supergravity \cite{das,frad}.
In this theory,
the rigid $SO(2)$ symmetry rotating the two independent
Majorana supersymmetries present in the ungauged theory, is made local
by  the introduction of a minimal gauge coupling between the graviphoton and
the gravitinos. Local supersymmetry then requires a negative cosmological
constant and a gravitino mass term.
The theory has four bosonic and four fermionic degrees of freedom;
it describes a graviton $V_{\mu}^{\alpha}$,
two Majorana gravitinos
$\psi_{\mu}^I$ $(I=1,2)$, which we combine into a single complex spinor
$\psi_{\mu} = \psi_{\mu}^1 + i\psi_{\mu}^2$,
and a Maxwell gauge field $A_{\mu}$, minimally
coupled to the gravitinos, with coupling constant $g$. 
The bosonic part of the Lagrangian is \cite{das,frad}
\eq
V^{-1}{\cal L} = -\frac{1}{4}R - \frac{1}{4}F_{\mu\nu}F^{\mu\nu}
+ \frac{3}{2}g^2, \label{lagrangeN2}
\feq
where $R$ is the scalar curvature,
$F_{\mu\nu} = \partial_{\mu}A_{\nu} - \partial_{\nu}A_{\mu}$
is the gauge field strength
and the cosmological constant is $\Lambda = 3g^2$.
We look for solutions of the field equations  from
(\ref{lagrangeN2}), with a product metric
\eq
ds^2 = f(r)dt^2 - f(r)^{-1}dr^2 - K^2d\Omega^2, \label{metric}
\feq
where $K$  is constant, and $d\Omega^2$ is a metric
of curvature $k=0,\pm1$ on a two-dimensional manifold $\Sigma$. 
For the gauge field $A$ we make the ansatz
\eq
A = \left\{ \begin{array}{ll}
\frac{q_e}{K^2} r\,dt + q_m\cos\theta d\phi & k=1 \\
\frac{q_e}{K^2} r\,dt + q_m\theta d\phi &k=0 \\
\frac{q_e}{K^2} r\,dt + q_m\cosh\theta d\phi & k=-1,
\end{array} \right. \label{ansgaugef}
\feq
$q_e$ and $q_m$ denoting the electric and magnetic charges 
respectively.
One easily checks that (\ref{ansgaugef}) solves the gauge field
equations of motion, $\partial_{\mu}(VF^{\mu\nu})=0$.
The equations of motion for the metric are
\eqn
&&\frac{d^2f(r)}{dr^2} = 2(Q^2 + \Lambda), \label{secderiv}\\
&&K^2(Q^2 - \Lambda) = k, \label{curv}
\feqn
where we have defined
\eq
Q^2 \equiv \frac{q_e^2 + q_m^2}{ K^4}.
\feq
From (\ref{secderiv}) we have
\eq
f(r)= (Q^2 + \Lambda)r^2.
\feq

In the following we are interested in bosonic backgrounds preserving
some amount of supersymmetry, which means that the gravitino variation
must vanish,
\eq
\de \psi_{\mu} = \hat{\nabla}_{\mu} \epsilon = 0.
\label{gravvar}
\feq
Here $\epsilon$ is an infinitesimal Dirac spinor, and $\hat{\nabla}_{\mu}$
is given by
\eq
\hat{\nabla}_{\mu} = D_{\mu} + \frac{i}{2}g\ga_{\mu} -
\frac{1}{2}F_{\al\be}\sigma^{\al\be}\ga_{\mu},
\feq
with the Lorentz and gauge covariant derivative
\eq
D_{\mu} = \partial_{\mu} + \frac{1}{2}\om_{\mu}^{\;\;\al\be}\sigma_{\al\be}
          +igA_{\mu}. \label{Lorgaugecovder}
\feq
We use standard conventions, $\{ \gamma_{\mu},\gamma_{\nu}\} =2g_{\mu\nu}$,
$\sigma_{\alpha\beta}=1/4[\gamma_{\alpha},\gamma_{\beta}]$.
The integrability conditions for (\ref{gravvar}) lead to the solution
\eq
q_e = 0, \quad q_m = \pm gK^2. \label{charges}
\feq
Inserting (\ref{charges}) into (\ref{curv}) yields for the curvature $k$ of
the two-manifold $\Sigma$
\eq
k = - 2g^2K^2 < 0, \label{curvsusy}
\feq
Thus $\Sigma$ must be diffeomorphic to the hyperbolic space
$H^2$ or to a quotient thereof. Without loss of generality setting
$k=-1$ we obtain  
\eq
K^2 =\frac{1}{ 2g^2}, \qquad q_m = \pm \frac{1}{2g}.
\feq
Introducing the
dimensionless coordinates $\tau = gt$, $\rho = gr$, 
the metric becomes
\eq
ds^2 = \frac{1}{2g^2}(8\rho^2 d\tau^2 - \frac{d\rho^2}{2\rho^2} -
       d\theta^2 - \sinh^2\theta d\phi^2). \label{metricsusy}
\feq
The $(\tau,\rho)$-part of the metric is just the line element of 
$AdS_2$ in horospherical coordinates.

A short comment on the supersymmetry conditions on the charges
is in order. Usually one has electromagnetic
duality invariance, i.e. electric and magnetic charges
enter  into the BPS conditions in a symmetric way.
In our case however, due to the minimal coupling of
the graviphoton to the gravitino,
this duality invariance is broken.
The bosonic sector remains duality invariant upon gauging, but
the Killing spinor equation does not, since the gauge potential $A_{\mu}$
appears in (\ref{Lorgaugecovder}). A similar nonsymmetric appearance
of electric and magnetic charges in the BPS conditions was found for
black holes in diverse gauged supergravity theories
\cite{cald,klemm,duff,sabra}.

Let us now turn back to the question of how many supersymmetries are
preserved by our solution. In general, the integrability conditions
are necessary, but not sufficient to guarantee the existence
of Killing spinors. Solving explicitly the
Killing spinor equations,
we find the solution
\eq
\epsilon(\tau,\rho) = \left[\sqrt{\rho}(1-\ga_1i) + (-4\tau\sqrt{\rho}\ga_0i
+ \frac{1}{\sqrt{\rho}})(1+\ga_1i)\right]P_{\pm}\epsilon_0,
\label{killing}
\feq
where $\epsilon_0$ denotes a constant spinor and
\eq
P_{\pm} \equiv \frac{1}{2}(1 \pm i\ga_2\ga_3)
\feq
is a projection operator . The $\pm$ sign corresponds to the sign
of the magnetic charge
$q_m = \pm1/(2g)$. We choose the $+$ sign for definiteness.
The appearance of the projector $P_{+}$ explicitly shows that
half of the ${\cal N}=2$ supersymmetries are broken; the dimension of the 
solution space is reduced from four to two (complex) dimensions.

In order to determine the residual symmetry superalgebra
of the above supergravity configurations we make
use of a technique described in \cite{gauntlett,town} (cf.~also
\cite{figueroa}). It is based
on the fact that, up to purely bosonic factors, the isometry superalgebra
is determined by the Killing spinors, just as the bosonic symmetry algebra
is determined by the Killing vectors. To see this, one first observes
that, given two Killing spinors $\epsilon$ and $\epsilon'$, the
bilinear $\bar{\epsilon}\ga^{\mu}\epsilon'\partial_{\mu}$ is a Killing
vector. In \cite{gauntlett}
it was shown that
\eq
\{Q_F(\epsilon), Q_F(\epsilon')\} = Q_B(\bar{\epsilon}\ga^{\mu}\epsilon'
                                    \partial_{\mu})
\feq
for the corresponding charges. This
means that the determination of the linear combination
$\bar{\epsilon}\ga^{\mu}\epsilon'\partial_{\mu}$ of Killing
vectors is equivalent
to the determination of the linear combination of bosonic charges
appearing in the anticommutator of any pair of fermionic charges.\\
We define
\eq
\eta_{\pm} = \frac{1}{2}(1 \pm i\gamma_1)P_+\epsilon_0. \label{eta}
\feq
Here $\frac{1}{2}(1 \pm i\gamma_1)$ is an additional projector that
commutes with $P_+$.
Using the $\eta_{\pm}$, we can write for the Killing spinors in
(\ref{killing})
\eq
\epsilon(\tau,\rho) = 2\sqrt{\rho}\eta_- + 2(-4\tau\sqrt{\rho}\ga_0i +
\frac{1}{\sqrt{\rho}})\eta_+ = \epsilon_-
+ \epsilon_+.
\feq
Now the Killing vectors 
$\xi_{++}$, $\xi_{+-}$, $\xi_{--}$ can be expressed in terms of
these Killing spinors as
$\xi_{++} = \bar{\epsilon}_+
\ga^{\mu}\epsilon_+\partial_{\mu}$, etc.
In this way we obtain
\eq
\xi_{++} = 32g(\bar{\eta}_+\ga_0\eta_+)\ell_-, \qquad \xi_{+-} =
-8ig(\bar{\eta}_+\eta_-)\ell_0, \qquad \xi_{--} = 2g(\bar{\eta}_-\ga_0\eta_-)
\ell_+,
\feq
where
\eqn
\ell_- &=& (\tau^2 + \frac{1}{16\rho^2})\partial_{\tau} - 2\tau\rho
\partial_{\rho}, \nonumber \\
\ell_+ &=& \partial_{\tau}, \nonumber\\
\ell_0&=& -\tau\partial_{\tau} + \rho\partial_{\rho}
\feqn
are the Killing vectors of $AdS_2$, satisfying the $so(2,1)$ algebra
\eqn
[\ell_+,\ell_-] &=& -2\ell_0, \nonumber \\
{[}\ell_0,\ell_{\pm}] &=& \pm \ell_{\pm}.
\feqn
We see that the anticommutator of two supercharges contains only
the $so(2,1)$ generators of $AdS_2$; the additional bosonic
$so(2,1)$ symmetries of hyperbolic space $H^2$ are not obtained by the
above method. This means that the isometry superalgebra of this
$AdS_2\times H^2$ background is given by a direct sum of an appropriate
superextension of $so(2,1) \cong su(1,1)$, and a bosonic part
$so(2,1) \cong su(1,1)$. In our case, we thus obtain
$osp(2|2)\oplus so(2,1) \cong su(1,1|1)\oplus su(1,1)$ for the residual
superalgebra.

Finally we want to exhibit a BPS solitonic
object  which interpolates between the above ${\cal N}=1$ supersymmetric
$AdS_2\times H^2$ solution and the maximally supersymmetric
$AdS_4$ spacetime. In fact such a soliton has been
found in \cite{cald}: it represents an extremal black hole with
metric
\eq
ds^2 = \left(rg - \frac{1}{2gr}\right)^2 dt^2 - \left(rg -
       \frac{1}{2gr}\right)^{-2}dr^2
       - r^2(d\theta^2 + \sinh^2\theta d\phi^2), 
       \label{metricbh}
\feq
and gauge field
\eq
A = q_m\cosh\theta d\phi, \quad q_m = \frac{1}{2g}.
    \label{gaugefield}
\feq
The spacetime with metric (\ref{metricbh}) has an event horizon at
$r = r_+ = 1/(g\sqrt{2})$. Introducing the new coordinates
$\rho = g(r - r_+)$, $\tau = gt$, one verifies that the near-horizon
limit of (\ref{metricbh}) is indeed the metric in (\ref{metricsusy}).
On the other hand, for large $r$, (\ref{metricbh}) gives
\eq
ds^2 = (-1+r^2g^2)dt^2 - (-1+r^2g^2)^{-1}dr^2 - r^2(d\theta^2 + \sinh^2\theta
       d\phi^2),
\feq
which is simply $AdS_4$ seen by an accelerated
observer \cite{vanzo}. The Killing spinors for the configuration
(\ref{metricbh}), (\ref{gaugefield}) have been determined
in \cite{cald}
\eq
\epsilon(r) = \left(rg - \frac{1}{2gr}\right)^{\frac{1}{4}} 
              (1-i\gamma_1) P_+ \epsilon_0.
\feq
Here $\epsilon_0$ is subject to a double projection, which reduces
the complex dimension of the solution space from four to one.
Near the horizon, where the black hole metric approaches the one in
(\ref{metricsusy}), we have a supersymmetry enhancement resulting in
a doubling of the Killing spinors. Using the same technique as above,
one finds for the black hole a residual superalgebra
$s(2)\oplus su(1,1)$, where $s(2)$
denotes the superalgebra introduced by Witten to formulate
supersymmetric quantum mechanics \cite{witten2}.

\section{Particle motion near the horizon} \label{confmech}

In this section we study the motion of a particle
with mass $m$ and magnetic charge $q$
in the near-horizon regime of the extremal BPS black holes discussed
above and find that it is governed by a model of conformal
mechanics \cite{claus,kallosh}\footnote{For further aspects of
the connection between black holes (0-branes) and conformal mechanics
cf.~\cite{gibbtown,youm1,youm2,behrndt,blum,michelson}.}.

We consider the $AdS_2 \times H^2$ solution of gauged
${\cal N}=2$, $D=4$ supergravity: the metric is given in (\ref{metricsusy}),
and the gauge field is $A = q_m\cosh\theta d\phi$, with 
magnetic charge $q_m = 1/(2g)$. 
Defining new coordinates
\eqn
&& \zeta =\frac{2\tau}{g}, \qquad \qquad\qquad\qquad~~
\qquad\qquad x=\frac{1}{g \sqrt{\rho}},\nonumber \\
&&\xi =\frac{1}{2g}\frac{\sinh \theta \sin \phi }{\cosh \theta +
\sinh \theta \cos \phi}, \qquad \qquad z=\frac{1}{g \sqrt{\cosh \theta +
\sinh \theta \cos \phi}},
\feqn
we obtain\footnote{$\tilde{A}$ is equal to $A$ up to a
gauge transformation, so in the following we will omit the tilde.}:
\eqn
&&ds^2 = \frac{d\zeta ^2}{g^4 x^4} -\frac{dx^2}{g^2 x^2}-
2 \left(\frac{d\xi ^2}{g^4 z^4} +\frac{dz^2}{g^2 z^2} \right), \nonumber\\
&&\tilde{A}=  \frac{1}{g^2 z^2}d \xi.
\label{metricagain}
\feqn
We use a Hamiltonian formalism and define
\eq
{\cal{H}}= g^{\mu \nu} (\Pi_{\mu}-qA_{\mu}) (\Pi_{\nu}-qA_{\nu}),
\feq
where $\Pi_{\mu}$ denote generalized momenta and $g_{\mu \nu}$  is the
metric. For our configuration this leads to
\eq
{\cal{H}}=g^4 x^4 \Pi_{\zeta}^2 -g^2 x^2 \Pi_{x}^2 -\frac{g^2 }{2}
z^2 \Pi_z ^2 -\frac{g^4 z^4}{2} \left( \Pi_{\xi} -\frac{q}{g^2 z^2}
\right)^2. \label{eq:hamilton}
\feq
The Hamilton equations are
\eqn
\dot{\zeta} &=& 2g^4 x^4 \Pi_{\zeta}, \nonumber \\
\dot{x} &=& -2g^2 x^2 \Pi_x, \nonumber \\
\dot{z} &=& -g^2 z^2 \Pi_z, \\
\dot{\xi} &=& -g^4 z^4 \left( \Pi_{\xi}-\frac{q}{g^2 z^2} \right), \nonumber
\feqn
and
\eqn
\dot{\Pi}_{\zeta} &=& 0, \nonumber \\
\dot{\Pi}_x &=& -4g^4 x^3 \Pi_{\zeta}^2 +2g^2 x \Pi_x ^2, \nonumber \\
\dot{\Pi}_{z} &=& g^2 z \Pi_z ^2 +2g^2 z \Pi_{\xi} (g^2 z^2 \Pi_{\xi}-q),\\
\dot{\Pi}_{\xi} &=& 0, \nonumber
\feqn
where the dot denotes the derivative with respect to an affine parameter 
$\lambda$.  Since the coordinates  $\zeta$ and $\xi$ are cyclic,
the associated conjugate
momenta $\Pi_{\zeta}$ and $\Pi_{\xi}$ are conserved. Another constant of motion
is given by
\eq
\frac{g^2 }{2}
z^2 \Pi_z ^2 +\frac{g^4 z^4}{2} \left( \Pi_{\xi} -\frac{q}{g^2 z^2}
\right) ^2 \equiv \frac{c^2}{2}.
\label{eq:costante}
\feq
Moreover we have the on-shell relation ${\cal{H}}=m^2$ where
$m$ is the mass of the particle. Thus we can identify 
$H=\Pi_{\zeta}$ with the Hamiltonian of the magnetic particle 
$(m,q)$,
\eq
H=\frac{1}{g^2 x^2}
\sqrt{g^2 x^2 \Pi_x ^2 +\frac{g^2 }{2}
z^2 \Pi_z ^2 +\frac{g^4 z^4}{2} \left( \Pi_{\xi} -\frac{q}{g^2 z^2}
\right) ^2 +m^2}.  \label{eq:emiltoniana}
\feq
In particular, if we impose (\ref{eq:costante}),
$H$ becomes the Hamiltonian for a particle
in one dimension, being $x$ and $\Pi_x$ the conjugate variables. 
Defining
\eq
u=x\Pi_x, \qquad p=\Pi_x,
\feq
and
\eq
f(u) = \frac{g^2 u^2}{2 \sqrt{g ^2 u^2 +\frac{c^2}{2}+m^2 }},
       \label{ffunctgauged}
\feq
the reduced Hamiltonian becomes
\eq
H = \frac{p^2}{2f(u)}.
    \label{scaleinv}
\feq
Thus we have obtained a conformal theory in $0+1$ dimensions.
The generators of the conformal algebra $so(2,1)$ are
\eq
D=\frac{1}{2}u, \qquad\qquad
H=\frac{p^2}{2f(u)},\qquad\qquad
K=\frac{1}{2}x^2 f(u),
\label{gen}
\feq
where $D$ is the generator of dilatations and $K$ the generator of
proper conformal transformations. They satisfy the Poisson bracket
algebra
\eq
\left[ D,H \right]_{PB} = H, \qquad \qquad
\left[ D,K \right]_{PB} = -K,  \qquad \qquad
\left[ H,K \right]_{PB} = 2D.
\feq
A copy of this algebra is obtained by considering the isometries on
the hyperbolic plane $H^2$.
In fact from (\ref{eq:costante}) we have
\eq
\Pi_{\xi}=\frac{\Pi_z ^2}{2\psi (v)},
\feq
where
\eq
v= z\Pi_z,\qquad\qquad\qquad
\psi (v)= 2g^2 v^2 \frac{q+ \sqrt{c^2 -g^2 v^2}}{q^2 -c^2 +g^2 v^2}.
\feq
Correspondingly we introduce 
\eq
\tilde{D}=\frac{1}{2}v, \qquad \qquad
\tilde{H}=\Pi_{\xi}=\frac{\Pi_z ^2}{2\psi(v)}, \qquad \qquad
\tilde{K}=\frac{1}{2}z^2 \psi(v),
\label{newgen}
\feq
which again satisfy the $so(2,1)$ algebra
\eq
\left[ \tilde{D},\tilde{H} \right]_{PB} = \tilde{H}, \qquad\qquad
\left[ \tilde{D},\tilde{K} \right]_{PB} = -\tilde{K}, \qquad\qquad
\left[ \tilde{H},\tilde{K} \right]_{PB} = 2\tilde{D}.
\feq
The generators in (\ref{gen}) and the ones in (\ref{newgen}) commute and
give rise to the $so(2,1) \times so(2,1)$ algebra.
In the same way as in \cite{kumar}, this symmetry can be extended to two copies
of the Virasoro algebra, where the generators are given by
\eqn
L_n &=& -\frac{i}{2}x^{1+n}p^{1-n}f^n, \nonumber \\
\tilde{L}_n &=& -\frac{i}{2}z^{1+n}\Pi_z^{1-n}\psi^n.
\feqn
Now we go back to the Hamiltonian (\ref{scaleinv}) and discuss its
symmetries in detail; in particular we will see where the above Virasoro
symmetry comes from.

\section{Symmetries of conformal mechanics} \label{symm}

The Hamiltonian in (\ref{scaleinv}) describes the most general 
scale-invariant system of one dynamical variable $q$ and canonical conjugate 
momentum $p$ \cite{kumar}, $f$ denoting an arbitrary function of $u = qp$. 

For example, Hamiltonians of this
type where shown to govern the dynamics of a particle
in the near-horizon region of an extremal Reissner-Nordstr\"om 
black hole \cite{claus}.
Also, the conformal mechanics model of De Alfaro, Fubini
and Furlan \cite{dff}, with Hamiltonian
\eq
H = \frac{p^2}{2} + \frac{g}{2q^2}, \label{dffmodel}
\feq
where $g$ is a dimensionless coupling constant, can be recovered from
(\ref{scaleinv}) by setting
\eq
f(u) = \frac{1}{1 + gu^{-2}}.
       \label{f}
\feq

In general one can show \cite{kumar} that such theories exhibit a Virasoro
symmetry with generators 
\eq
L_n = -\frac{i}{2}q^{1+n}p^{1-n}f^n, 
\label{virasoro}
\feq
obeying 
\eq
[L_n,L_m]_{PB} = -i(m-n)L_{m+n}.
                 \label{viralg}
\feq
In particular the $so(2,1)$ subalgebra
(the conformal algebra in 0+1 dimensions) is generated by
\eq
iL_{-1}= H, \qquad\quad iL_0 = D,\qquad \quad
iL_1 = K,
\feq
with $H$, $D$ and $K$ as in (\ref{gen}). 
Let us now examine in more detail
why this Virasoro symmetry arises, and how it generalizes to $w_{\infty}$.
We consider the action of a particle in 0+1 dimensions,
\eq
S = \int dt (p\dot{q} - H(p,q)) = \int \alpha,
    \label{wact}
\feq
$\alpha$ being the one-form
\eq
\alpha = p\,dq - H\,dt.
\feq
A symmetry transformation for the action in (\ref{wact})
must leave invariant the two-form
$\Omega = d\alpha = dp \wedge dq - dH \wedge dt$ \cite{witten}. 
In order to determine these symmetries 
explicitly, first we perform the canonical transformation
\eq
q' = q\sqrt f, \qquad \quad p' = \frac{p}{\sqrt f}.
     \label{cantransf}
\feq
In terms of the new variables the Hamiltonian becomes
\eq
\tilde{H} = H = \frac{p'^2}{2}, \label{transfHam}
\feq
showing that classically the system is equivalent to a free particle.
Setting
\eq
y = p', \qquad\quad x = q' - p't, \label{2ndtransf}
\feq
the two-form $\Omega$ reduces to
\eq
\Omega = dy \wedge dx.
         \label{Omega}
\feq
Therefore the symmetries of the system are the diffeomorphisms
which preserve the symplectic two-form (\ref{Omega}). They are generated
by vector fields (cf.~\cite{sezgin})
\eqn
\xi &=& \xi^a\partial_a = \Omega^{ab}(\partial_b \Lambda(y,x) +
        \omega_b)\partial_a + h(y,x,t)\partial_t \nonumber \\
    &=& \frac{\partial \Lambda}{\partial y}\partial_{x} -
        \frac{\partial \Lambda}{\partial x}\partial_{y} + \omega_{y}
        \partial_{x} - \omega_{x}\partial_{y} + h\partial_t,
        \label{vecfields}
\feqn
where $\Lambda(y,x)$ and $h(y,x,t)$ are arbitrary functions, and
$\omega \in H^1(M,\R)$, $M$ denoting the $(x,y)$ phase space. This
symmetry algebra is isomorphic to that of the matrix model with
the standard inverted harmonic oscillator Hamiltonian \cite{aj1,minic,witten}.
In order to establish a correspondence with the symmetries arising
in string theories with two-dimensional target space \cite{aj1,minic,witten},
one should
instead consider only time-independent functions $h$ \cite{witten},
i.~e.~$h = h(y,x)$. In this case, the transformations generated by
(\ref{vecfields}) preserve $\Omega$ as well as the volume form
\eq
\Theta = dy \wedge dx \wedge dt.
\feq
For $h=0$,
the vector fields (\ref{vecfields}) generate the algebra $w_{\infty}$ of
area-preserving (symplectic) diffeomorphisms \cite{sezgin} on which
we will concentrate in the following\footnote{Note
that, in order for $w_{\infty}$ to admit central extensions, the first
cohomology group $H^1(M)$ ($M$ denoting $(x,y)$ phase space)
must be non-trivial. There exist $b_1$
independent central extensions, where $b_1$ is the first Betti number
of $M$ \cite{sezgin}.}.
Explicitly one can choose as a basis of generators 
\eq
v^l_m = y^{l+1}x^{l+m+1} = \left(\frac{p}{\sqrt f}\right)^{l+1}(q\sqrt f
- \frac{p}{\sqrt f}t)^{l+m+1},
 \label{basis} 
\feq
where $l = 0,1,\ldots$ and $m \in \Z$.
They satisfy  the $w_{\infty}$ algebra
\eq
[v^l_m,v^{l'}_{m'}]_{PB} = [m(l'+1) - m'(l+1)]v^{l+l'}_{m+m'}.
\label{winf}
\feq
The Virasoro algebra in (\ref{viralg}) is the one generated by
the $v^{-n}_{2n}$. 

One easily checks that
\eq
\frac{dv^l_m}{dt} = \frac{\partial v^l_m}{\partial t} + [v^l_m,H] = 0,
\feq
so the $v^l_m$ are constants of motion (conserved charges). The two fundamental
integrals of motion are $y = p/\sqrt f$ and $x = q\sqrt f - pt/\sqrt f$.

Now, whenever
\eq
f(u)\stackrel{u\rightarrow 0}{\rightarrow}
\left(\alpha+\frac{\beta}{u^2}\right)^{-1}
\label{asympt}
\feq
the spectrum of the Hamiltonian in (\ref{scaleinv}) 
is continuous and bounded from below ($E > 0$), the ground state at $E = 0$
being non-normalizable.\footnote{We observe that the functions $f(u)$ in
(\ref{ffunctgauged}) and in (\ref{f}) as well as the one of
the model in \cite{claus}
all satisfy the asymptotic behaviour in (\ref{asympt}).}.
A way to cure this
infrared problem is to study the time evolution of the system by
means of a compact operator \cite{dff} 
(which has a discrete set of normalizable eigenfunctions)
\eq
R = \frac12 \left(aH + \frac 1a K\right), \label{R}
\feq
where $a$ denotes an infrared cutoff. 
To be more specific we choose
\eq
K = q^2f/2.
\label{compact}
\feq 
Though scale and
translational invariance are broken by this procedure, the new model
admits a symmetry algebra isomorphic to $so(2,1)$ \cite{dff}.
In fact the new system is again invariant under the algebra $w_{\infty}$. 
Let us consider
$R$, with $H$ given by (\ref{scaleinv}) and $K$ as in (\ref{compact}).
Now we have
\eq
\Omega = dp \wedge dq - dR \wedge dt,
\feq
which, by means of the (canonical) transformation
\eqn
\bar{p} &=& \frac{1}{\sqrt 2} e^{-\frac{it}{2} }(\frac{ap}{\sqrt f}+iq\sqrt f),
\nonumber \\
\bar{q} &=& \frac{i}{a\sqrt 2}e^{\frac{it}{2}} (\frac{ap}{\sqrt f}- iq\sqrt f)
\label{cantransf2}
\feqn
can be recast into the form
\eq
\Omega = d\bar{p} \wedge d\bar{q}.
\feq
Similarly to the previous case, we take as basis functions
\eq
v^l_m = \bar{p}^{l+1}\bar{q}^{l+m+1}, \label{basis2}
\feq
which satisfy (\ref{winf}). As above, the $v^l_m$ are conserved,
the fundamental integrals of motion being $\bar{p}$ and $\bar{q}$.
A Virasoro subalgebra is spanned by
\eq
L_m=\frac{1}{4a} e^{imt} (\frac{ap}{\sqrt f} +iq\sqrt f)^{1-m}
(\frac{ap}{\sqrt f} -iq\sqrt f)^{1+m}.
\label{gen2}
\feq
In contrast to the generators in (\ref{virasoro}) we have now $L_{-m}=L^{\dag}$
and $L_0=R$, which means that $L_0$ plays the role of the Hamiltonian.

Let us finally discuss the question of central
charges. 
Since the Virasoro symmetry of this model
stems from the larger symmetry algebra of area-preserving diffeomorphisms, 
central extensions might be possible.
As already observed, in order to
have central extensions, the first Betti number $b_1$ of the manifold on which
these diffeomorphisms act, must be nonzero. Now we show that this is
indeed the case. To this end we go back to the Hamiltonian $R$ in (\ref{R})
with $H$ and $K$ given in (\ref{scaleinv}) and (\ref{compact})
respectively.
First of all we note that the $w_{\infty}$ generators (\ref{basis2}) are
singular at $\bar{q}=0$ for $m$ sufficiently negative. In order to
define smooth diffeomorphisms we
perform a canonical transformation
\eq
\varphi = 2\arctan\frac{qf}{ap},\qquad \qquad p_{\varphi} = \frac{ap^2}{4f} +
\frac{q^2 f}{4a}.
\feq
This yields the transformed Hamiltonian
\eq
\tilde{R} = R = p_{\varphi},
\feq
so classically the system is equivalent to a relativistic particle on the
circle. (Note that $0 \le \varphi \le 2\pi$, in order to have $q \ge 0$).
The two-form $\Omega$ becomes
\eq
\Omega = dp_{\varphi} \wedge d\varphi',
\feq
where $\varphi' = \varphi - t$. Thus, for constant $t$, the diffeomorphisms
preserving $\Omega$ are the area-preserving diffeomorphisms
SDiff($S^1 \times \R$) on the cylinder. This manifold has $b_1 = 1$, therefore
SDiff($S^1 \times \R$) admits exactly one central extension \cite{sezgin}.

\section{Conformal mechanics and bulk asymptotic symmetries}\label{adscft}

In this section we  show that the Virasoro generators (\ref{gen2})
of the infrared-regularized model with Hamiltonian $R$ given by
(\ref{R}) can be associated in a natural way to the generators of
the asymptotic symmetries of $AdS_2$ \cite{cadoni}. 

Let us first consider the $AdS_2$ metric with coordinates as
in (\ref{metricagain})
\eq
ds^2 =\frac{1}{g^4q^4}dt^2 -\frac{1}{g^2q^2}dq^2.
\feq
Then the asymptotic symmetries
of $AdS_2$ determined in \cite{cadoni} become
\eq
l_m = e^{imt}\left\{\left[1 - \frac{m^2g^2q^4}{8} + o(q^8)\right]\partial_t
+ \left[\frac{im}{2}q + o(q^2)\right]\partial_q\right\}. 
\label{asymptsymm}
\feq
They generate the diffeomorphisms which preserve the
asymptotic form\footnote{Note that the boundary is at $q=0$.}
of the $AdS_2$ metric. In \cite{cadoni}, the generators (\ref{asymptsymm})
have been determined by imposing certain boundary conditions on the
metric. These boundary conditions must be weak enough in order to allow
for a larger symmetry algebra than $so(2,1)$, but strong enough to
ensure the possibility to define the associated conserved charges
\cite{cadoni}. It is easy to show that the bulk generators (\ref{asymptsymm})
satisfy the Virasoro algebra
\eq
[l_m,l_k] = -i(m-k)l_{m+k}.
\feq
They can be rewritten as
\eq
l_m = e^{imt}\left\{-(m^2-1)l_0 + \frac m2 (m+1)e^{-it}l_1
+ \frac m2 (m-1)e^{it}l_{-1}\right\} \label{asymptsymm1}
\feq
In order to establish the correspondence with the symmetries of
our conformal mechanics model we associate the generators $l_0,l_{\pm 1}$ 
with the phase-space functions given in (\ref{gen2})
\eqn
&&l_0\rightarrow L_0 = R, \nonumber \\
&&l_{\pm1}\rightarrow L_{\pm 1} = e^{\pm it}\left\{R - \frac{q^2f}{2a} \mp
\frac i2 qp\right\}.
\label{correspondence}
\feqn
In this way, using (\ref{correspondence}) in
(\ref{asymptsymm1}), we find 
\eq
l_m\rightarrow \tilde{L}_m = e^{imt}\left\{\frac{ap^2}{4f} + \frac{q^2f}{4a} -
\frac{m^2}{2a}q^2f - \frac{im}{2}qp\right\}.
\label{gencorr}
\feq
At this point it is trivial to check that,
when expanded
in a Laurent series in $q$ near the boundary $q=0$, 
the $\tilde{L}_m$ in (\ref{gencorr}) obtained 
through the above correspondence and the $L_m$ of the conformal
mechanics given in (\ref{gen2}) do agree up to the
order $q^6$ inclusive. The IR-regularized 
model realizes explicitly the asymptotic symmetries of
$AdS_2$.

The central extension of the algebra is easily 
obtained through a comparison with the work in ref.~\cite{cadoni}.
There the asymptotic symmetries of $AdS_2$ were realized canonically
in the Hamiltonian formulation of the Jackiw-Teitelboim (JT) model.
In this way, the authors of \cite{cadoni} found a
central charge, which is is expressed in terms of
the dilaton field evaluated at the black hole horizon. In our case,
in order to identify the dilaton we proceed as follows:
we consider static solutions of the form
\eqn
&&ds^2= h_{ij}(r,t) dx^i dx^j-\eta^2(r,t) \sigma_{IJ}dx^I dx^J, \nonumber \\
&&A=A_t(r)dt+A_{\phi}(\theta)d\phi,
\label{spheransatz}
\feqn
where $i,j=t,r$; $I,J=\theta,\phi$, and $\sigma_{IJ}$ denotes the metric on
a two-dimensional space of constant curvature $k$.
From Gauss theorem we have
\eq
F_{tr}=q_e \frac{\sqrt{-h}}{\eta^2}, \qquad\quad 
F_{\theta\phi} =q_m \sqrt{\sigma},
\label{emfield}
\feq
where 
\eq
\sqrt{\sigma} = \sqrt{\det{\sigma_{IJ}}} = \left\{
\begin{array}{ll}
\sin\theta & k=1,\\
\theta & k=0,\\
\sinh\theta & k=-1.
\end{array}
\right.
\feq
The ansatz for the metric and the electromagnetic field strength
can be used in the Lagrangian (\ref{lagrangeN2}) of ${\cal N}=2$
gauged supergravity. The corresponding equations of motion give rise to an
effective two-dimensional dilatonic theory of gravity,
being $h_{ij}$ the
metric and $\eta$ the dilaton field. The action becomes
\eq
S = \Omega\int dx\,dt \sqrt{-h} \left[-\frac 12 (\nabla \eta)^2
    -\frac 14 \eta^2 R + V(\eta)\right],
\feq
where the dilaton potential is given by
\eq
V(\eta) = \frac k2 - \frac{q_e^2 + q_m^2}{2\eta^2} + \frac{3g^2\eta^2}{2},
          \label{dilpot}
\feq
and $\Omega$ denotes the volume of the reduction
space\footnote{In the case of noncompact spaces, one should consider
a quotient thereof, i.~e.~a Riemann surface of genus $n \ge 1$.}.\\
We consider now generic black holes with metric
\eq
ds^2 = W(r)dt^2 - W(r)^{-1}dr^2 - r^2 d\Omega^2,
\feq
where
\eq
W(r) = k - \frac{2m}{r} + g^2r^2 + \frac{q_e^2+q_m^2}{r^2}.
\feq
Near the horizon $r=r_+$, where $W(r)=0$, the metric contains an $AdS_2$
factor if the black hole is extremal. This leads to the requirement
\eq
W(r_+) = 2V(r_+),
\feq
where $V(r)$ is the dilaton potential (\ref{dilpot}).
This means that at the horizon, the dilaton potential
(\ref{dilpot}) vanishes. Near the horizon, we expand
\eq
\eta^2(r) = (r_+ + \Phi)^2 \to r_+^2 + 2r_+\Phi, \qquad \Phi \to 0,
\feq
$\Phi = 0$ denoting the horizon position. We further have
\eqn
(\nabla \eta)^2 &\to& \left(1 - \frac{2\Phi}{r_+}\right)(\nabla \Phi)^2,
                      \nonumber \\
V(\eta) &\to& V(\Phi) \equiv V'(r_+)\Phi.
\feqn
Putting the dilaton on-shell ($\Phi = r-r_+$), neglecting a Gauss-Bonnet
term, and defining
\eq
\bar{\Phi} = \frac 12 r_+ \Phi,
\feq
we arrive at the JT model with action
\eq
S = \Omega\int dx\,dt \sqrt{-h} \bar{\Phi}(-R + 2\lambda^2),
\feq
where
\eq
\lambda^2 = \frac 12 W''(r_+) = \frac{V'(r_+)}{r_+}.
\feq
At this point we can use the results of ref.~\cite{cadoni}
to obtain that for all the various superconformal particle
models the central charge is given by
\eq
c = 24\Omega \sqrt{\frac{r_+^3}{V'(r_+)}}.
\feq
In particular, for the interpolating soliton (\ref{metricbh}),
(\ref{gaugefield}), one gets
\eq
c = \frac{12\Omega}{g^2\sqrt{2}}. 
\feq

We emphasize that the results presented so far are quite general:
they are valid for any conformal quantum mechanics describing the motion of a 
particle in a black hole configuration that in the near-horizon region is
given by $AdS_2 \times \Sigma_2$ where $\Sigma_2$ is a two-dimensional 
manifold with constant curvature.

\section{Superextensions}\label{super}

We now ask whether one can find generalizations of conformal
mechanics which are invariant under superextensions of the $w_{\infty}$
algebra. Superconformal mechanics has first been studied in
\cite{akulov,fubini}, where the $su(1,1|1) \cong osp(2|2)$
generalization of the DFF model was constructed. The authors of \cite{claus}
studied the motion of a superparticle in the $AdS_2 \times S^2$
background of an extreme RN black hole in the near-horizon limit.
There, an $osp(1|2)$
superextension of the DFF model was considered, which, however, was
only possible for $g=0$ in (\ref{dffmodel}), if one wants the
standard supersymmetry to be linearly realized (cf.~also \cite{aziz}).
This is not surprising,
as for $g \neq 0$ there is no classical solution of zero energy, so
there can be no ground state annihilated by the supercharge $Q$ \cite{claus}.
Below we shall construct models of ${\cal  N} =1$ or ${\cal  N} =2$
superconformal mechanics in which
both the standard supersymmetry and the conformal supersymmetry are
nonlinearly realized. These models will turn out to admit not only the
$osp(1|2)$ or $osp(2|2)$ symmetry algebras, but the entire ${\cal  N} =1$
or ${\cal  N} =2$
superextensions of $w_{\infty}$.

Let us first consider the ${\cal  N} =1$ case. We start
from the bosonic Hamiltonian
(\ref{scaleinv}), which, after the canonical transformation (\ref{cantransf})
reduces to (\ref{transfHam}), i.~e.~to the Hamiltonian of a free particle.
For a free particle, however, an ${\cal  N} =1$ superextension can
easily be found,
the action is given by
\eq
S = \int dt\,(\frac12 \dot{q}'^2 + \frac i2 \psi\dot{\psi}), \label{act1}
\feq
where $\psi$ is an anticommuting world line field. Performing now the
inverse of the canonical transformation (\ref{cantransf}), the action
(\ref{act1}) goes over in
\eq
S = \int dt\,(p\dot{q} - \frac{p^2}{2f} + \frac i2 \psi\dot{\psi})
\label{act2}
\feq
plus a surface term, which we can drop. We are thus led to propose
(\ref{act2}) as ${\cal  N} =1$ superparticle action, which generalizes the
DFF model for $g \neq 0$. In the following we will show that the
symmetry algebra admitted by (\ref{act2}) is indeed the ${\cal  N} =1$
superextension of $w_{\infty}$. Similar to the bosonic case,
(\ref{act2}) can be written as
\eq
S = \int \alpha,
\feq
where
\eq
\alpha = p\,dq + \frac i2 \psi\,d\psi - H\,dt,
\feq
so the two-form $\Omega = d\alpha$ considered in section \ref{symm} reads
\eq
\Omega = dp \wedge dq - dH \wedge dt + \frac i2 d\psi \wedge d\psi.
\feq
Using (\ref{cantransf}) and (\ref{2ndtransf}), one finds for the
super two-form
\eq
\Omega = dy \wedge dx + \frac i2 d\psi \wedge d\psi.
\feq
It is now easy to identify the vector fields which preserve $\Omega$;
they are of the form\footnote{We omitted possible harmonic one-forms $\omega$.}
\eqn
\xi &=& \Omega^{AB}\partial_B \Lambda \partial_A + h\partial_t \nonumber \\
&=& \frac{\partial \Lambda}{\partial y}\partial_{x} -
\frac{\partial \Lambda}{\partial x}\partial_{y} -
i\frac{\partial \Lambda}{\partial \psi}\partial_{\psi} + h\partial_t,
\label{xi}
\feqn
where $\Lambda = \Lambda(y,x,\psi)$ and $h = h(y,x,\psi,t)$ are
arbitrary superfunctions.
As a basis on the supermanifold parametrized by the coordinates
$y,x,\psi$, we can take
\eqn
v^l_m &=& y^{l-1}x^{l+m-1} = \left(\frac{p}{\sqrt f}\right)^{l-1}(q\sqrt f
- \frac{p}{\sqrt f}t)^{l+m-1}, \nonumber \\
G^l_r &=& y^{l-1}x^{l+r-1}\psi = \left(\frac{p}{\sqrt f}\right)^{l-1}
(q\sqrt f - \frac{p}{\sqrt f}t)^{l+r-1}\psi. \label{basisN=1}
\feqn
Under the super Poisson bracket \cite{sezsok,pope}
\eq
[f,g] = \frac{\partial f}{\partial q}\frac{\partial g}{\partial p} -
\frac{\partial f}{\partial p}\frac{\partial g}{\partial q} +
2\frac{\partial f}{\partial \psi}\frac{\partial g}{\partial \psi},
\label{superpoisson}
\feq
the basis functions (\ref{basisN=1}) generate the ${\cal  N}=1$ superextension
of $w_{\infty}$ \cite{sezstr89}, which reads
\eqn
[v^k_m,v^l_n] &=& [m(l-1) - n(k-1)]v^{k+l-2}_{m+n}, \nonumber \\
{[}G^k_r,G^l_s] &=& 2 v^{k+l-1}_{r+s}, \\
{[}v^k_m,G^l_r] &=& [m(l-1) - r(k-1)]G^{k+l-2}_{m+r}. \nonumber
\feqn
This is an algebra of symplectic super-diffeomorphisms \cite{sezsok}. Note
that the $v^l_m$ and the $G^l_r$ are conserved (super)charges. An
${\cal  N} =1$ super-Virasoro subalgebra is generated by
\eqn
L_n &=& -\frac{i}{2} v^{2-n}_{2n} = -\frac{i}{2}\left(\frac{p}
{\sqrt f}\right)^{1-n}
(q\sqrt f - \frac{p}{\sqrt f}t)^{1+n},
\nonumber \\
G_r &=& -\frac{i}{\sqrt 2} G^{3/2-r}_{2r} = -\frac{i}{\sqrt 2}
\left(\frac{p}{\sqrt f}
\right)^{1/2-r}(q\sqrt f - \frac{p}{\sqrt f}t)^{1/2+r}\psi,
\feqn
where $n \in \Z$ and $r \in \Z + \frac 12$ in the Neveu-Schwarz sector
and $r \in \Z$ in the Ramond sector. The $L_n$ and $G_r$ satisfy
\eqn
[L_m,L_n] &=& -i(m-n)L_{m+n}, \nonumber \\
{[}G_r,G_s] &=& -2iL_{r+s}, \\
{[}L_m,G_r] &=& -i(\frac m2 - r)G_{m+r}.
\feqn
This super-Virasoro algebra generalizes the bosonic part found in
 \cite{kumar}. It contains an $osp(1|2)$ subalgebra, whose generators
read
\eq
H = iL_{-1}, \quad D = iL_0, \quad K = iL_1, \quad Q = iG_{-1/2},
\quad S = iG_{1/2}. \label{osp12gen}
\feq
We now show that the supersymmetries are nonlinearly realized.
The infinitesimal variation of
a superfunction $F(y,x,\psi,t)$ under a symplectic superdiffeomorphism
generated by $\xi$ in (\ref{xi}), is given by
\eq
\delta_{\xi}F = {\cal L}_{\xi}F = \xi^A\partial_A F, \label{deltaF}
\feq
${\cal L}_{\xi}$ denoting the Lie derivative along $\xi$.
Taking $\Lambda = \epsilon Q$ and $h=0$ in (\ref{xi}), where $\epsilon$
is constant and anticommuting, and $Q$ was defined in (\ref{osp12gen}),
one gets
\eq
\delta_{\epsilon Q}q = \frac{\epsilon \psi}{\sqrt{2f}}(1 - \frac{pqf'(u)}{2f}),
\label{susyvarq}
\feq
which, for the DFF model (\ref{dffmodel}), reduces to $\delta_{\epsilon Q}q =
\sqrt{\frac f2}\epsilon \psi$, and
\eq
\delta_{\epsilon Q}\psi = -\frac{ip}{\sqrt{2f}}\epsilon.
\label{susyvarpsi}
\feq
Similar relations hold for $\Lambda = \epsilon S$, i.~e.~for the
variations under conformal supersymmetry transformations.
(\ref{susyvarq}) and (\ref{susyvarpsi}) show that the supersymmetries are
indeed nonlinearly realized.

One can also generalize the model considered above to ${\cal  N} =2$.
In this case,
the action reads ($\alpha,\beta = +,-$)
\eq
S = \int dt\,(p\dot{q} - \frac{p^2}{2f} + \frac i2 \psi^{\alpha}\delta_{\alpha
\beta}\dot{\psi}^{\beta}).
\feq
For the super two-form $\Omega$ we get
\eqn
\Omega &=& dp \wedge dq - dH \wedge dt + \frac i2 \delta_{\alpha\beta}
d\psi^{\alpha} \wedge d\psi^{\beta}, \nonumber \\
&=& dy \wedge dx + \frac i2 \delta_{\alpha\beta}
d\psi^{\alpha} \wedge d\psi^{\beta}
\feqn
by means of (\ref{cantransf}) and (\ref{2ndtransf}). It is preserved by
the vector fields
\eq
\xi = \frac{\partial \Lambda}{\partial y}\partial_{x} -
\frac{\partial \Lambda}{\partial x}\partial_{y} -
i\frac{\partial \Lambda}{\partial \psi^{\alpha}}
\partial_{\psi^{\alpha}} + h\partial_t,
\feq
with the superfunctions $\Lambda = \Lambda(y,x,\psi^{\alpha})$ and
$h = h(y,x,\psi^{\alpha},t)$. In super phase space we take the basis
\eqn
v^l_m &=& y^{l-1}x^{l+m-1} = \left(\frac{p}{\sqrt f}\right)^{l-1}(q\sqrt f
- \frac{p}{\sqrt f}t)^{l+m-1}, \nonumber \\
G^{l\pm}_r &=& y^{l-1}x^{l+r-1}\psi^{\pm} = \left(\frac{p}{\sqrt f}
\right)^{l-1}(q\sqrt f - \frac{p}{\sqrt f}t)^{l+r-1}\psi^{\pm}, \\
J^l_m &=& \frac 12 y^{l-2} x^{l+m-2}\psi^+\psi^- = \frac 12
\left(\frac{p}{\sqrt f}
\right)^l (q\sqrt f - \frac{p}{\sqrt f}t)^{l+m}\psi^+\psi^-.
\nonumber
\feqn
We define the graded Poisson bracket \cite{pope}
\eq
[f,g] = \frac{\partial f}{\partial q}\frac{\partial g}{\partial p} -
\frac{\partial f}{\partial p}\frac{\partial g}{\partial q} -
2(-1)^{\deg f}\left(\frac{\partial f}{\partial \psi^+}\frac{\partial g}
{\partial \psi^-} + \frac{\partial f}{\partial \psi^-}\frac{\partial g}
{\partial \psi^+}\right), \label{gradbrack}
\feq
where $\deg f$ is the grading of $f$. Using (\ref{gradbrack}), our basis
functions generate the ${\cal  N} =2$ superextension of
$w_{\infty}$ \cite{pope},
\eqn
[v^k_m,v^l_n] &=& [m(l-1) - n(k-1)]v^{k+l-2}_{m+n}, \nonumber \\
{[}G^{k-}_r,G^{l+}_s] &=& 2 v^{k+l-1}_{r+s} - 2[r(l-1) - s(k-1)]
                          J^{k+l-1}_{r+s}, \nonumber \\
{[}v^k_m,G^{l\pm}_r] &=& [m(l-1) - r(k-1)]G^{k+l-2,\pm}_{m+r}, \\
{[}v^k_m,J^l_n] &=& [m(l-2) - n(k-1)]J^{k+l-2}_{m+n}, \nonumber \\
{[}J^k_m,G^{l\pm}_r] &=& \pm G^{k+l-2,\pm}_{m+r}. \nonumber
\feqn
It contains an ${\cal  N} =2$ super-Virasoro subalgebra with generators
\eqn
L_n &=& -\frac{i}{2} v^{2-n}_{2n} = -\frac{i}{2} \left(\frac{p}
{\sqrt f}\right)^{1-n}
(q\sqrt f - \frac{p}{\sqrt f}t)^{1+n},
\nonumber \\
G_r^{\pm} &=& -\frac{i}{\sqrt 2} G^{3/2-r}_{2r} = -\frac{i}{\sqrt 2}
\left(\frac{p}{\sqrt f}\right)^{1/2-r}(q\sqrt f - \frac{p}
{\sqrt f}t)^{1/2+r}\psi^{\pm}, \\
J_n &=& -iJ^{2-n}_{2n} = -\frac{i}{2} \left(\frac{p}{\sqrt f}\right)^{-n}
(q\sqrt f - \frac{p}{\sqrt f}t)^n\psi^+\psi^-,
\nonumber 
\feqn
satisfying
\eqn
[L_m,L_n] &=& -i(m-n)L_{m+n}, \nonumber \\
{[}G_r^-,G_s^+] &=& -2iL_{r+s} - i(s-r)J_{r+s}, \nonumber \\
{[}L_m,G_r^{\pm}] &=& -i(\frac m2 - r)G_{m+r}^{\pm}, \\
{[}L_m,J_n] &=& in J_{m+n}, \nonumber \\
{[}J_m,G_r^{\pm}] &=& \pm iG_{m+r}^{\pm}. \nonumber
\feqn
Like above, we have $n \in \Z$ and $r \in \Z + \frac 12$ in the
Neveu-Schwarz sector and $r \in \Z$ in the Ramond sector. 
The $osp(2|2) \cong su(1,1|1)$ subalgebra is spanned by
\eq
H = iL_{-1}, \quad D = iL_0, \quad K = iL_1, \quad Q^{\pm} = iG_{-1/2}^{\pm},
\quad S^{\pm} = iG_{1/2}^{\pm}, \quad iJ_0. \label{osp22gen}
\feq

\section{Conclusions}\label{disc}

The initial aim of our work was the study of the (super)particle 
dynamics in a BPS black hole background solution of
${\cal N}=2$ gauged supergravity in $D=4$ dimensions. We have found that
such a system is described by a quantum mechanical model whose
symmetry group is generated by a couple of conformal algebras,
one in the angular sector and one in the time-radial sector.
The angular conformal symmetry is new, and occurs due to the hyperbolic
event horizon geometry. Restricting ourselves to a fixed value of the
angular $so(2,1)$ Casimir, we showed that the particle Hamiltonian reduces
to a universal scale invariant form, and that its symmetries extend to the
algebra $w_{\infty}$ of area-preserving diffeomorphisms, acting in
phase space.\\
Naively, the conformal Virasoro subalgebra of
$w_{\infty}$ has vanishing central charge \cite{kumar}. On the
other hand, we showed that a one-to-one correspondence can be established
between this Virasoro algebra and the asymptotic symmetries of $AdS_2$.
In addition in ref.~\cite{cadoni} it had been proven that these asymptotic
symmetries exhibit a nonvanishing central charge. The crucial observation
to resolve this puzzle was the fact that the Virasoro algebra
of the conformal mechanics model appears as a subalgebra of $w_{\infty}$
acting on a manifold with $b_1=1$, i.~e.~with exactly one possible central
extension. The central charge was then computed by dimensional reduction of
the bulk action, yielding a Jackiw-Teitelboim model, and comparison with the 
work in ref.~\cite{cadoni}. Thus our system offers an explicit realization
of the $AdS_2/CFT_1$ correspondence.

\section*{Acknowledgements}
The part of this work due to D.~K.~has been partially
supported by a research grant
within the {\em Common Special Academic Program III} of the
Federal Republic of Germany and its Federal States, mediated 
by the DAAD.\\
The authors would like to thank G.~Berrino, M.~Cadoni, M.~M.~Caldarelli,
V.~Moretti and L.~Vanzo for helpful discussions.

\end{document}